\documentclass[aps, prb, twocolumn, showpacs, superscriptaddress]{revtex4-1}

\usepackage{amsmath,amssymb}
\usepackage{graphicx}
\usepackage{dcolumn}
\usepackage{bm}
\begin{document}

\title {$Z_{2}$ fractionalized Chern/topological insulators in an exactly soluble correlated model}
\author{Yin Zhong}
\email{zhongy05@hotmail.com}
\affiliation{Center for Interdisciplinary Studies $\&$ Key Laboratory for
Magnetism and Magnetic Materials of the MoE, Lanzhou University, Lanzhou 730000, China}
\author{Yu-Feng Wang}
\affiliation{Center for Interdisciplinary Studies $\&$ Key Laboratory for
Magnetism and Magnetic Materials of the MoE, Lanzhou University, Lanzhou 730000, China}
\author{Hong-Gang Luo}
\email{luohg@lzu.edu.cn}
\affiliation{Center for Interdisciplinary Studies $\&$ Key Laboratory for
Magnetism and Magnetic Materials of the MoE, Lanzhou University, Lanzhou 730000, China}
\affiliation{Beijing Computational Science Research Center, Beijing 100084, China}

\date{\today}

\begin{abstract}
In this paper we propose an exactly soluble model in two-dimensional honeycomb lattice, from which two phases are found. One is the usual Chern/topological insulating state and the other is an interesting $Z_2$ fractionalized Chern/topological insulator. While their bulk properties are similar, the edge-states of physical electrons are quite different. The single electron excitation of the former shows a free particle-like behavior while the latter one is gapped, which provides a definite signature to identify the fractionalized states. The transition between these two phases is found to fall into the 3D Ising universal class. Significantly, near the quantum transition point the physical electron in the edge-states shows strong Luttinger liquid behavior. An extension to the interesting case of the square lattice is also made. In addition, we also discuss some relationship between our exactly soluble model and various Hubbard-like models existing in the literature. The essential difference between the proposed $Z_{2}$ fractionalized Chern insulator and the hotly pursued fractional Chern insulator is also pointed out. The present work may be helpful for further study on the fractionalized insulating phase and related novel correlated quantum phases.
\end{abstract}

\maketitle

\section{Introduction} \label{intr}
The topological insulators/superconductors discovered recently, which cannot be described
by the classic Landau symmetry-breaking theory,\cite{Wen2004,Sachdev2011,Continentino} have motivated many studies in these novel states of matter.\cite{Bernevig2006,Kane2006,Wiedmann,Shen2009,Hasan2010,Qi2011,Kitaev2008,
Schnyder2008,Ran2008,Qi2008,Levin2009,Rachel,Hohenadler2011,Ruegg2012,Yoshida,Zhong2013a}
The topological insulators are gapped in the bulk but have gapless helical edge-state protected by the time-reversal symmetry.
Recently, the idea of topological insulators has been expanded into the symmetry-protected topological (SPT) states, which are bulk-gapped quantum phases with symmetries, and have gapless or degenerate boundary states as long as the symmetries are not broken.\cite{Chen2011a,Chen2011b,Chen2013,Wen2012,Levin2012,Lu2012,Senthil2013,Vishwanath,Grover2013,Lu2012b,Wang2013}

Another kind of state of matter beyond the Landau symmetry-breaking paradigm is the well-known integer quantum Hall (IQH) states and the fractional quantum Hall (FQH) states.\cite{Wen2004}
The former exhibits integer quantized Hall conductance in the external magnetic field while the latter shows the topological order and has exotic fractional charge excitations and fractional statistics.
The Chern insulator is a band insulator exhibiting a nonzero quantized Hall conductance but preserving the lattice
translational symmetry. It is a natural extension of IQH in the lattice systems without any external magnetic field.\cite{Haldane,Regnault} When including strong interaction with appropriate electron-filling, fractional Chern insulator may exist in certain models and their nature has been extensively explored. \cite{Neupert,Mei,Sun,Regnault,Sheng2011,Wang2012a,Neupert2012,Nakamura,Bergholtz,Barkeshli2012x,Wu2012,Lu2012c,McGreevy}

In most cases, interaction effect in above mentioned topological states is not easy to treat, thus many approximate analytical and/or numerical techniques and phenomenological approaches have to be utilized to acquire intuitive physics. Here we show a kind of exactly soluble models which may realize certain fractionalized Chern/topological insulators-like states in suitable parameter space. Our proposed model is motivated by the study of the orthogonal metals,\cite{Nandkishore,Zhong2012e,Zhong2012,Zhong2013b} which have the same thermal and transport behaviors as the usual Landau Fermi liquid but with gapped single particle spectrum. Authors in Refs.[\onlinecite{Nandkishore,Zhong2012e,Zhong2012,Zhong2013b}]
mostly focused on the issue of the exotic metallic states with the aim at possible non-Fermi liquid behaviors and the elusive critical Fermi surfaces.\cite{Senthil2008} Here we are interested in possible $Z_{2}$ fractionalized states, which are hard to be identified by their bulk properties but their edge-states of physical electrons are gapped in contrast to their non-fractionalized counterparts, which can be used to identify these $Z_{2}$ fractionalized states. This is an interesting feature shown in the present paper, which has not been reported in the previous works.

The model we proposed is defined on various two dimensional lattices at half-filling with the superficial $Z_{2}$ gauge structure. We provide an explicit formalism for the case of the honeycomb lattice and extend it to the interesting square lattice case. Using the dual transformation for the $Z_{2}$ lattice gauge fields, the original model can be written as two decoupling Hamiltonian. One part describes a quantum transverse Ising model while the other is just free auxiliary fermion similar to the celebrated Haldane or Kane-Mele model.\cite{Haldane,Kane2006} The free auxiliary fermion always contributes quantized charge or spin Hall conductance while the dual quantum Ising model supports a second-order quantum phase transition. The behavior of physical electrons is determined by combining these two parts and we find that the bulk properties are difficult to use for distinguishing $Z_{2}$ fractionalized Chern/topological insulators from the usual Chern/topological insulating states.

Significantly, the edge-states of physical electrons have rather different behavior for the fractionalized and the usual non-fractionalized states, which provides a definite signature to identify the fractionalized states from the non-fractionalized ones. In the fractionalized states, since physical electrons lose their coherence, the corresponding single electron excitation in edge-states are gapped while gapless edge-states survive in the usual Chern/topological insulating states because the auxiliary free fermions have non-zero weight of the physical electrons. (We should emphasize that although the single electron excitation is gapped in the fractionalized states, the static and dynamical many-particle correlations, e.g. density-density or spin-spin correlation, are still gapless in those states as what have been shown in Ref.[\onlinecite{Ruegg2012}].)\cite{Ruegg2013c} More interesting, when the dual Ising model approaches its quantum critical point, the Green's function of physical electron in the edge-states will show the strong Luttinger liquid behavior in contrast to the case of usual free particles.

In addition, we have further made a comparable study to the fractionalized quantum spin Hall (QSH$^{\ast}$) state in Ref.\onlinecite{Ruegg2012} and have extended our discussion to the case of the square lattice with a modified model. Particularly, for the square lattice, we expect that a chiral topological superconducting phase and its $Z_{2}$ fractionalized version may appear when the attractive local interaction is introduced. Since our exactly soluble models may not appear in many real models for condensed matter physics, in term of the $Z_{2}$ slave-spin mean-field approximation,\cite{deMedici,Ruegg} we have suggested that similar fractionalized states may appear in various Hubbard-like models without intrinsic degree of freedom for gauge fields. The relation of $Z_{2}$ fractionalized Chern insulators found in the present work to the fractional Chern insulator Refs. [\onlinecite{Neupert,Mei,Sun,Regnault,Sheng2011}] is explored, which shows that such two kinds of fractionalized states are rather distinct in their nature and constructing exactly soluble models for the desirable fractionalized Chern insulators is still widely open.\cite{Nakamura} We also provide a brief discussion on the relation to orthogonal metals\cite{Nandkishore,Zhong2012e,Zhong2012,Zhong2013b} and fractional topological insulators.\cite{Levin2009}

The remainder of this paper is organized as follows. In Sec. \ref{sec1}, we first introduce the basic exactly soluble model without spin degree of freedom on the honeycomb lattice and discuss two useful limit cases for this model. Then Sec. \ref{sec2} provides the solution of the basic model in terms of a dual transformation for the $Z_{2}$ gauge-field. States with and without $Z_{2}$ fractionalization are found and one of them is identified as the $Z_{2}$ fractionalized Chern insulator state while the other is the usual Chern insulator. The bulk features are distinct in the topological level, which emphasizes the confinement and deconfinement of two states. Properties of the corresponding edge-states are studied in detail. In Sec. \ref{sec3}, the spin degree of freedom is included and the resulting states are also analyzed. In Sec. \ref{sec4}, we present some possible extensions including the interesting soluble model on the square lattice and discussions on related issues. Finally, Sec. \ref{sec5} is devoted to a brief conclusion.
\section{an exactly soluble model}\label{sec1}
We propose following model defined on the honeycomb lattice at half-filling,
\begin{eqnarray}
&&H=H_{I}+H_{c},\nonumber\\
&&H_{I}=-h\sum_{\langle ij\rangle}\hat{\sigma}^{z}_{ij}-J\sum_{i}(-1)^{c^{\dag}_{i}c_{i}}\prod_{j=i+\delta_{a}}\hat{\sigma}^{x}_{ij}\nonumber\\
&&-W\sum_{i}\prod_{j\in hexagon}\sigma^{z}_{ij},\nonumber\\
&&H_{c}=-t\sum_{\langle ij\rangle}c_{i}^{\dag}\hat{\sigma}^{z}_{ij}c_{j}-t'\sum_{\langle\langle ij\rangle\rangle}e^{i\varphi_{ij}}c_{i}^{\dag}\hat{\sigma}^{z}_{il}\hat{\sigma}^{z}_{lj}c_{j},\label{eq1}
\end{eqnarray}
where $H_{I}$ describes a modified $Z_{2}$ lattice gauge theory model and $H_{c}$ denotes the coupling of conduction electrons $c_{i}$ to the former $Z_{2}$ field $\hat{\sigma}^{z}_{ij}$ (Ising field). (The $Z_{2}$ field $\hat{\sigma}^{z}_{ij},\hat{\sigma}^{x}_{ij}$ are usual Pauli matrices whose commutation relation is $[\hat{\sigma}^{\alpha}_{ij},\hat{\sigma}^{\beta}_{i'j'}]=2i\epsilon_{\alpha\beta\gamma}\delta_{ii'}\delta_{jj'}\hat{\sigma}^{\gamma}_{ij}$. In other words, $Z_{2}$ field in different sites commutes with each other while the ones on the same sites obey the usual commutation relation of spin-1/2 Pauli matrix.) In $H_{I}$, $h$-term can
be considered as the kinetic energy while $J$-term acts like a potential ($j=i+\delta_{a}$ denotes three nearest-neighbor sites). [Readers who are not familiar with the lattice gauge theory may find Ref.[\onlinecite{Kogut}] readable and useful.] $H_{c}$ will be the standard Haldane model defined on the honeycomb lattice if $Z_{2}$ gauge field is ignored. The first term of $H_{c}$ is the usual hopping term between nearest-neighbor sites with tuning by the $Z_{2}$ field $\hat{\sigma}^{z}_{ij}$. The phase $\varphi_{ij}=\pm\frac{1}{2}\pi$ in the next-nearest-neighbor hopping term in $H_{c}$ is introduced to give rise to a quantized Hall conductance without external magnetic fields and the positive phase is gained with anticlockwise hopping. Here two $Z_{2}$ gauge fields are introduced to enforce the $Z_{2}$ gauge invariance and the label $l$ in $\hat{\sigma}^{z}_{il}\hat{\sigma}^{z}_{lj}$ should be taken as the intermediate site between $i$ and $j$ sites.

One may find our model is similar to the exactly soluble models provided in the study of the orthogonal metals.\cite{Nandkishore,Zhong2012e,Zhong2012,Zhong2013b} Actually, this model is motivated by those works but we here focus on different respects.
We are interested in the possible unusual gapped states while those authors mostly studied the metallic states in order to find possible non-Fermi liquid behaviors and the exotic critical Fermi surfaces.
Since this model is insulating, its ground-state is hard to be identified by their bulk properties but one will find that the edge-states of physical electrons play a crucial role. This is not studied in the previous works for the orthogonal metals and is an interesting feature in the present paper. In addition, the authors in Ref.[\onlinecite{Ruegg2012}] also studied the unusual $Z_{2}$ gapped states (QSH$^{\ast}$) with the $Z_{2}$ slave-spin mean-field approach, we will discuss this point in Sec. \ref{sec3} and \ref{sec4}.

Below we first analyze two simplified cases, which recover two models extensively discussed in the literature.

\subsection{The modified $Z_{2}$ gauge theory model on the honeycomb lattice}
We first consider the first part $H_I$ decoupled with the conduction electrons, the resultant Hamiltonian reads
\begin{eqnarray}
H_{I}&&=-h\sum_{\langle ij\rangle}\hat{\sigma}^{z}_{ij}-J\sum_{i}\prod_{j=i+\delta_{a}}\hat{\sigma}^{x}_{ij}\nonumber\\
&&-W\sum_{i}\prod_{j\in hexagon}\sigma^{z}_{ij}.\label{eq2}
\end{eqnarray}
A careful reader may find this model could be considered as a Kitaev toric code model defined on a honeycomb lattice with a external field term $h$.\cite{Kitaev} Therefore, readers who are familiar with that model can skip this subsection without missing any important physics. This model has the $Z_{2}$ gauge symmetry (structure) and can be seen as follows. A careful reader may find this model could be considered as a Kitaev toric code model defined on a honeycomb lattice.
First, one can define the $Z_{2}$ gauge transformation operator as $\hat{G}_{i}=\prod_{j\in hexagon}\sigma^{z}_{ij}$ and this operator flips all Ising gauge field $\hat{\sigma}^{x}_{ij}$ in the $i$-th hexagon. Since $J$-term covers two Ising gauge fields in the same hexagon, this term is obviously unchanged when the $Z_{2}$ gauge transformation $\hat{G}_{i}$ is utilized. Therefore, the Hamiltonian (\ref{eq2}) is invariant under the operation of $\hat{G}_{i}$, thus it has the expected $Z_{2}$ gauge symmetry (invariance). One can also check that $[\hat{G}_{i},H_{I}]=0$, which means that the Hamiltonian has the local (gauge) symmetry enforced by $\hat{G}_{i}$. Since $\hat{G}_{i}^{2}=1$, one may choose $\hat{G}_{i}=\pm1$. Moreover, the physical states of the Hamiltonian should be invariant under the operation of $\hat{G}_{i}$. (For a brief discussion on the standard $Z_{2}$ gauge theory model one can refer to Appendix A.)

There are two limit situations which are easy to analyze. First, if $J,W\gg h$, the ground-state should have $\prod_{j=i+\delta_{a}}\hat{\sigma}^{x}_{ij}=1$ and this is just the deconfined state for the $Z_{2}$ gauge theory. However, when $J,W\ll h$, the ground-state can be obtained by setting all $\hat{\sigma}^{z}_{ij}=1$ and the resultant state is the confined state for the $Z_{2}$ gauge theory. More importantly, any physical excitation must carry zero gauge charge in the confined state in contrast to the case of deconfined state where fractionalized excitation with nonzero gauge charge are permitted in principle. Such feature for the excitation will be manifested when matter fields (electrons) are included in the gauge theory model.

To sharpen our understanding of the above model, it is useful to carry out a dual transformation which transforms the $Z_{2}$ lattice gauge theory model into a simple (quantum) Ising model living on the honeycomb lattice. The dual transformation is defined as
\begin{equation}
\prod_{j=i+\delta_{a}}\hat{\sigma}^{x}_{ij}=\hat{\tau}^{z}_{i}, \hat{\sigma}^{z}_{ij}=\hat{\tau}^{x}_{i}\hat{\tau}^{x}_{j}.\nonumber
\end{equation}
Then, the dual Hamiltonian reads as
\begin{eqnarray}
H_{I}=-h\sum_{\langle ij\rangle}\hat{\tau}^{x}_{i}\hat{\tau}^{x}_{j}-J\sum_{i}\hat{\tau}^{z}_{i},\label{eq3}
\end{eqnarray}
which is just the well-known quantum transverse Ising model on the honeycomb lattice and the $W$ term is neglected since it only contributes constant energy.\cite{Sachdev2011} We should emphasize that the $Z_{2}$ gauge transformation operator $\hat{G}_{i}=\prod_{j=i+\delta_{a}}\hat{\sigma}^{x}_{ij}=1$ due to above dual transformation and no constraints are needed for the dual Hamiltonian in contrast to the original one. The ground-state phase diagram of the quantum transverse Ising model is clear: it has two phases characterized by the usual Landau local order parameter $\langle \hat{\tau}^{x}_{i} \rangle=0$ (the paramagnetic phase) and $\langle\hat{\tau}^{x}_{i}\rangle\neq0$ (the ferromagnetic phase).\cite{Sachdev2011} Besides, a second-order quantum phase transition exists between those two phases whose critical behaviors belong to the 3D Ising universal class.\cite{Sachdev2011}

Some readers may wonder why we do not use the standard $Z_{2}$ gauge theory model on the honeycomb lattice, which is more friendly and is more easy to analyze its topological order (For a detail see Appendix A). The reason is that
in the standard model, the $Z_{2}$ gauge transformation operator $\hat{G}_{i}$ cannot automatically be unit after the dual transformation and this leads to unnecessary extra complications. Therefore, we use the modified $Z_{2}$ gauge theory model which reflects the same physics without caring about the unsatisfactory constraints. However, the modified $Z_{2}$ gauge theory model is not friendly if one considers its topological order although the modified $Z_{2}$ gauge theory model itself indeed shows the same topological order as the standard one.

Additionally, if one is only interested in the low-energy physics, the following $\varphi^{4}$ theory may be useful in this respect,
\begin{equation}
Z=\int D\phi e^{-\int d\tau d^{2}x [(\partial_{\tau}\phi)^{2}+c^{2}(\nabla\phi)^{2}+r\phi^{2}+u\phi^{4}]},\label{eq4}
\end{equation}
where $r,u$ are effective parameters depending on microscopic details. (For the derivation and brief discussion on this effective action one can refer to Appendix B.)

According to the above dual transformation, it is readily to see that the paramagnetic state of the quantum Ising model corresponds to the deconfined state of the original $Z_{2}$ gauge theory model while the ferromagnetic phase dualizes to the confined state. Moreover, the quantum critical point of the quantum Ising model translates to the deconfinement-confinement transition point though no Landau local order parameter can be reliably defined in the case of gauge theory model. (The Wilson loop can be defined in the gauge theory model but it is not a local operator and we will not study this point further in our present paper.)

\subsection{The pure Haldane model}
In this subsection, we will briefly review some important properties of the pure Haldane model (without any gauge fields coupled to the electrons), which is useful for our next discussion.

Our starting point is the usual Haldane model on the honeycomb lattice at half-filling \cite{Haldane}
\begin{eqnarray}
H_{H}=-t\sum_{\langle ij\rangle }c_{i}^{\dag}c_{j}-t'\sum_{\langle\langle ij\rangle\rangle}e^{i\varphi_{ij}}c_{i}^{\dag}c_{j}. \label{eq5}
\end{eqnarray}
It is useful to rewrite this single-particle Hamiltonian in the momentum space as
\begin{eqnarray}
&&H_{H}=-t\sum_{k}[f(k)c_{kA}^{\dag}c_{kB}+f^{\star}(k)c_{kB}^{\dag}c_{kA}] \nonumber\\
&& \hspace{2cm} + 2t'\gamma(k)(c_{kA}^{\dag}c_{kA}-c_{kB}^{\dag}c_{kB}), \nonumber
\end{eqnarray}
where we have defined $f(k)=e^{-ik_{x}}+2e^{ik_{x}/2}\cos(\frac{\sqrt{3}}{2}k_{y})$, $\gamma(k)=\sin(\sqrt{3}k_{y})-2\cos(\frac{3}{2}k_{x})\sin(\frac{\sqrt{3}}{2}k_{y})$ and $A$, $B$ representing two nonequivalent sublattices of the honeycomb lattice, respectively. Then, by diagonalizing the above Hamiltonian, one
obtains the quasiparticle energy band as
\begin{eqnarray}
E_{k\pm}=\pm\sqrt{t^{2}|f(k)|^{2}+4t'^{2}\gamma(k)^{2}}, \nonumber
\end{eqnarray}
which preserves the particle-hole symmetry. It is well-known that for $3\sqrt{3}t'<t$, the excitation gap mainly opens near six Dirac points (Only two of them are nonequivalent in fact).\cite{Rachel} Then, expanding both $f(k)$ and $\gamma(k)$ near two nonequivalent Dirac points $\pm\vec{K}=\pm(0,\frac{4\pi}{3\sqrt{3}})$, respectively, the gap can be found as $\Delta_{gap}=6\sqrt{3}t'$ and the quasiparticle energy reads $E_{q\sigma\pm}\simeq\pm\sqrt{(\frac{3}{2}tq)^{2}+(3\sqrt{3}t')^{2}}$ with $q=(q_{x},q_{y})\equiv(k_{x},k_{y}\mp\frac{4\pi}{3\sqrt{3}})$.

The most interesting property of the Haldane model is that it gives rise to a quantized Hall conductance without applying the external magnetic field in contrast to the usual quantum Hall effect. The corresponding quantized Hall conductance can be readily calculated by $\sigma_{H}=C_{1}\frac{e^{2}}{h}$ where $C_{1}=\frac{1}{4\pi}\int dk_{x}dk_{y}\hat{\textbf{d}}\cdot(\frac{\partial\hat{\textbf{d}}}{\partial k_{x}}\times \frac{\partial\hat{\textbf{d}}}{\partial k_{y}})$.\cite{Qi2011} For the Haldane model, the single particle Hamiltonian can be rewritten as $\hat{h}(k)=\textbf{d}(k)\cdot \hat{\bf{\sigma}}$ with $\textbf{d}=(-t \text{Re} f(k),t \text{Im} f(k),2t'\gamma(k))$, $\hat{\textbf{d}}=\textbf{d}/|\textbf{d}|$ and $\hat{\bf{\sigma}}$ being the usual Pauli matrices. Then, by inserting the expression of $\hat{\textbf{d}}$ into the formula of $C_{1}$, one obtains $C_{1}=1$ and $\sigma_{H}=\frac{e^{2}}{h}$
for the Haldane model. In literature, the ground-state of the Haldane model is called the Chern insulator since it
exhibits a quantized Hall conductance without external magnetic fields and preserves the lattice translational symmetry of the honeycomb lattice.\cite{Haldane,Regnault} Hereafter, we will use the Chern insulator when involving the ground-state for the Haldane model.

\subsection{The low-energy effective theory for the pure Haldane model}
For the discussion of the low-energy physics, an effective $2+1D$ massive Dirac action can be obtained by expanding original Haldane model around two nonequivalent Dirac points $\pm\vec{K}$,
\begin{eqnarray}
S_{H}=\int d^{2}xd\tau \mathcal{L}_{0}=\int d^{2}xd\tau\sum_{a}[\bar{\psi}_{a}(\gamma_{\mu}\partial_{\mu}+m)\psi_{a}], \nonumber
\end{eqnarray}
where $\gamma_{\mu}=(\tau_{z},\tau_{x},\tau_{y})$ and $\partial_{\mu}=(\partial_{\tau},\partial_{x},\partial_{y})$
with $\tau_{z},\tau_{x},\tau_{y}$ the Pauli matrices. Here the same indices mean summation. We have introduced the effective mass $m=-3\sqrt{3}t'$ of Dirac fermions and set the effective Fermi velocity $v_{F}=\frac{3}{2}t$ to unit. The Dirac fields are defined as $\psi_{1\sigma}=(c_{1A\sigma},c_{1B\sigma})^{T}$, $\psi_{2\sigma}=(c_{2A\sigma},-c_{2B\sigma})^{T}$ and $\bar{\psi}_{a\sigma}=\psi^{\dag}_{a\sigma}\gamma_{0}$ with $a=1,2$ denoting the states near the two nonequivalent Dirac points $\pm\vec{K}=\pm(0,\frac{4\pi}{3\sqrt{3}})$ and $T$ implying the transposition manipulation.

If the external electromagnetic field $A_{\mu}=(i\phi,A_{x},A_{y})$ is introduced by the conventional minimal coupling ($\partial_{\mu}\rightarrow\partial_{\mu}-ieA_{\mu}$), the resulting effective Dirac action coupled to the external electromagnetic field reads
\begin{eqnarray}
S=\int d^{2}xd\tau\sum_{a}[\bar{\psi}_{a}(\gamma_{\mu}(\partial_{\mu}-ieA_{\mu})+m)\psi_{a}].  \label{eq6}
\end{eqnarray}
By integrating out the Dirac fields, we get an effective Chern-Simons action, which represents the nontrivial electromagnetic response of the massive Dirac fermions to the external electromagnetic field $A_{\mu}$, \cite{He2011}(For details, see Appendix C.)
\begin{eqnarray}
S_{CS}=\int d^{2}xd\tau[Ne^{2}\frac{-im}{8\pi|m|}\epsilon^{\mu\nu\lambda}A_{\mu}\partial_{\nu}A_{\lambda}],  \label{eq7}
\end{eqnarray}
where $N=2$ (from the two nonequivalent Dirac points), $\epsilon^{\mu\nu\lambda}$ is the usual all-antisymmetric tensor and the regular Maxwell term ($\sim F^{2}_{\mu\nu}$) has been dropped out since the low energy physics is dominated by the Chern-Simons term alone. Then, the physically observable quantized Hall conductance can be obtained from $J_{x}=\frac{\partial S_{CS}}{\partial A_{x}}|_{\vec{A}\rightarrow0}=Ne^{2}\frac{-im}{4\pi|m|}(\partial_{y}A_{0}-\partial_{0}A_{y})=\frac{e^{2}}{2\pi}E_{y}$ and we reproduce the result for the Hall conductance $\sigma_{H}=\frac{e^{2}}{h}$
where $h=2\pi\hbar$ is reintroduced with $m/|m|=-1$ and $N=2$.\cite{Wen2004}

Alternatively, one may inspect the low-energy physics from the perspective of the corresponding edge-state. Firstly,
the following effective Chern-Simon action reproduces the quantized Hall conductance of the Haldane model
\begin{equation}
S=\int d^{2}xdt[\frac{1}{4\pi}\epsilon^{\mu\nu\lambda}a_{\mu}\partial_{\nu}a_{\lambda}+\frac{e}{2\pi}\epsilon^{\mu\nu\lambda}A_{\mu}\partial_{\nu}a_{\lambda}],\label{eq8}
\end{equation}
where $a_{\mu}$ is the auxiliary dynamic gauge field and it gives rise to the physical charge/particle current $j_{\mu}=\frac{1}{2\pi}\epsilon^{\mu\nu\lambda}\partial_{\nu}a_{\lambda}$. According to the formalism in quantum Hall effect,\cite{Wen2004} the so-called K-matrix in the present action is $K=1$ with the charge-vector $q=1$, thus the physical quantized Hall conductance is calculated as $\sigma_{H}=qK^{-1}q\frac{e^{2}}{h}=\frac{e^{2}}{h}$. Moreover, the ground-state of the present model is not degenerated in the torus due to $|K|=1$ and the elementary quasiparticle is the usual fermion (electron) since the the exchange statistical angle $\theta$ is $\pi$, which means that the exchange of two identical quasiparticle leads to a $\pi$ phase acquired in their wavefunction.
The corresponding edge-state can be easily derived by standard bulk-edge correspondence for the effective abelian Chern-Simons theory,\cite{Wen2004}
\begin{eqnarray}
S_{edge}&&=\int dxdt\frac{1}{4\pi}[\partial_{t}\phi\partial_{x}\phi-v(\partial_{x}\phi)^{2}]\nonumber\\
&&+\frac{1}{2\pi}(\partial_{t}\phi A_{x}-\partial_{x}\phi A_{t})\label{eq9}
\end{eqnarray}
with $v$ denoting the non-universal velocity of edge states and $\phi$ being the bosonic representation for the edge-state modes. This edge-state can be refermionized by introducing the fermion operator $\psi\propto e^{i\phi}$ and the resulting action reads
\begin{eqnarray}
S_{edge}=\int dx dt[\psi^{\dag}(i\partial_{t}+eA_{t}-iv\partial_{x}-veA_{x})\psi].\label{eq10}
\end{eqnarray}
The above edge-state mode contributes $e^{2}/h$ to the quantized Hall conductance and its gaplessness or the exactly quantized value for the Hall conductance is protected by the chiral feature of the edge-state.

\section{The Chern insulator and the $Z_{2}$ fractionalized Chern insulator} \label{sec2}
Having studied the pure Haldane Eq.(\ref{eq5}) and the $Z_{2}$ gauge theory model Eq.(\ref{eq2}), it is ready to discussion the full model Eq.(\ref{eq1}) with both gauge field (Ising field) and matter field (electrons). For convenience of discussion, here we rewrite Eq. (\ref{eq1}) as follows
\begin{eqnarray}
&&H=H_{I}+H_{c},\nonumber\\
&&H_{I}=-h\sum_{\langle ij\rangle}\hat{\sigma}^{z}_{ij}-J\sum_{i}(-1)^{c^{\dag}_{i}c_{i}}\prod_{j=i+\delta_{a}}\hat{\sigma}^{x}_{ij}\nonumber\\
&&-W\sum_{i}\prod_{j\in hexagon}\sigma^{z}_{ij},\nonumber\\
&&H_{c}=-t\sum_{\langle ij\rangle}c_{i}^{\dag}\hat{\sigma}^{z}_{ij}c_{j}-t'\sum_{\langle\langle ij\rangle\rangle}e^{i\varphi_{ij}}c_{i}^{\dag}\hat{\sigma}^{z}_{il}\hat{\sigma}^{z}_{lj}c_{j}.\nonumber
\end{eqnarray}
Comparing this model to the $Z_{2}$ gauge theory model Eq.(\ref{eq2}) and its related $Z_{2}$ gauge transformation operator, one may use the same
$\hat{G}_{i}=\prod_{j\in hexagon}\sigma^{z}_{ij}$, which leads to the happy result $[\hat{G}_{i},H]=0$. Thus, the full model Eq.(\ref{eq1}) have the desirable $Z_{2}$ gauge structure. One may note that the physical electrons do not carry any gauge charge of the Ising field $\hat{\sigma}^{x}_{ij}$, which means even though the Ising field $\hat{\sigma}^{x}_{ij}$ appears in its confined state, the electrons are still well-defined physical excitations in this case.

Here, let us first inspect the model (\ref{eq1}) without resorting to its dual formalism. Since the physical electrons are gapped due to the next-nearest-neighbor hopping, one may integrate them out and the resulting Hamiltonian will be like the effective pure gauge field part\cite{Ruegg2012}
\begin{eqnarray}
\tilde{H}_{I}=-\widetilde{h}\sum_{\langle ij\rangle}\hat{\sigma}^{z}_{ij}-\widetilde{J}\sum_{i}\prod_{j=i+\delta_{a}}\hat{\sigma}^{x}_{ij}-\widetilde{W}\sum_{i}\prod_{j\in hexagon}\sigma^{z}_{ij}.\nonumber
\end{eqnarray}

If the $Z_{2}$ gauge-field is in its deconfined state and $\widetilde{W}$ is small, one may set all $\prod_{j=i+\delta_{a}}\hat{\sigma}^{x}_{ij}=1$ and neglect other two terms. Then the low-energy excitation in this case is the so-called
$Z_{2}$ charge.\cite{Ruegg2012} The $Z_{2}$ charge near site $i$ can be seen as a manifold with $\prod_{j=i+\delta_{a}}\hat{\sigma}^{x}_{ij}=-1$ with others unchanged. When $\widetilde{W}$ becomes large, the system remains in the deconfined phase but there exists another low-lying excitation the $Z_{2}$ vortex (vison), which is created by setting certain $\prod_{j\in hexagon}\sigma^{z}_{ij}=-1$ while keeping others intact.\cite{Senthil2000} One can see that $\widetilde{h}$-term adds or destroys the $Z_{2}$ charge ($\hat{\sigma}^{x}_{ij}$ in $\widetilde{h}$-term flip the spin on the link thus creates or destroys the $Z_{2}$ charge.) while $\widetilde{W}$-term do not change the number of $Z_{2}$ charge ($\widetilde{W}$-term commutes with $\widetilde{J}$-term.). In addition, the deconfined $Z_{2}$ gauge theory has the so-called topological order, (A characteristic signature of the topological order is
the ground-state degeneracy depending on the topology of the system) e.g., there exists four (two) degenerate ground-states if the system is put on a torus (cylinder).\cite{Senthil2000b}

We also note that an effective mutual $U(1)\times U(1)$ Chern-Simons theory can describe the low-energy behavior of the $Z_{2}$ gauge theory model.\cite{Kou2008}
\begin{eqnarray}
S=\int d^{2}xd\tau\left[\frac{1}{\pi}\epsilon^{\mu\nu\lambda}a_{\mu}\partial_{\nu}b_{\lambda}+a_{\mu}j_{\mu}+b_{\mu}J_{\mu}\right]\nonumber
\end{eqnarray}
where $j_{\mu}$ and $J_{\mu}$ represent the $Z_{2}$ charge and the $Z_{2}$ vortex, respectively. $a_{\mu}$, $b_{\mu}$ are the auxiliary gauge fields whose Chern-Simons term represents the semionic
mutual statistics between the $Z_{2}$ charge and the $Z_{2}$ vortex. Since the $K$-matrix is $K=\left(
                                                                                                  \begin{array}{cc}
                                                                                                    0 & 2 \\
                                                                                                    2 & 0 \\
                                                                                                  \end{array}
                                                                                                \right)
$ and $|DetK|=4$, the mutual $U(1)\times U(1)$ Chern-Simons theory correctly reproduces the four degenerate ground-state on a torus.

When the gauge-field is confined, the $\widetilde{h}$-term will dominate and the number of $Z_{2}$ charge is not well-defined. In other words, the $Z_{2}$ charge is condensed in this situation and we may consider the confined state is
just the condensed state for the $Z_{2}$ charge. The ground-state in the large $\widetilde{h}$ limit is obtained by setting all $\hat{\sigma}^{z}_{ij}=1$. Instead, the elementary excitation is the $Z_{2}$ link created by $\sigma^{z}_{ij}=-1$. In addition, the previous mutual $U(1)\times U(1)$ Chern-Simons theory is not useful in this state due to the confined nature.

Then, we will immerse into the discuss with the help of the dual transformation. Following the same treatment in the last section, the dual transformation is defined by $\prod_{j=i+\delta_{a}}\hat{\sigma}^{x}_{ij}=\hat{\tau}^{z}_{i}, \hat{\sigma}^{z}_{ij}=\hat{\tau}^{x}_{i}\hat{\tau}^{x}_{j}$ and the resulting Hamiltonian reads
\begin{eqnarray}
&&H=H_{I}+H_{c}\nonumber,\\
&&H_{I}=-h\sum_{\langle ij\rangle}\hat{\tau}^{x}_{i}\hat{\tau}^{x}_{j}-J\sum_{i}(-1)^{c^{\dag}_{i}c_{i}}\hat{\tau}^{z}_{i},\nonumber\\
&&H_{c}=-t\sum_{\langle ij\rangle}c_{i}^{\dag}\hat{\tau}^{x}_{i}\hat{\tau}^{x}_{j}c_{j}-t'\sum_{\langle\langle ij\rangle\rangle}e^{i\varphi_{ij}}c_{i}^{\dag}\hat{\tau}^{x}_{i}\hat{\tau}^{x}_{j}c_{j}. \label{eq11}
\end{eqnarray}
The above Hamiltonian can be further transformed via $f_{i}\equiv\hat{\tau}^{x}_{i}c_{i},(-1)^{c^{\dag}_{i}c_{i}}\hat{\tau}^{z}_{i}\rightarrow\hat{\tau}^{z}_{i}$ and one arrives at a simple formalism
\begin{eqnarray}
&&H=H_{\tau}+H_{f},\nonumber\\
&&H_{\tau}=-h\sum_{\langle ij\rangle}\hat{\tau}^{x}_{i}\hat{\tau}^{x}_{j}-J\sum_{i}\hat{\tau}^{z}_{i},\nonumber\\
&&H_{f}=-t\sum_{\langle ij\rangle}f_{i}^{\dag}f_{j}-t'\sum_{\langle\langle ij\rangle\rangle}e^{i\varphi_{ij}}f_{i}^{\dag}f_{j}. \label{eq12}
\end{eqnarray}
Comparing to Eq.[\ref{eq5}], one can see that $H_{f}$ is identical to the usual Haldane model (Eq.[\ref{eq5}]) except that the $f$ fermion is not the original electron $c$ but acts like a slave-particle. $H_{\tau}$ is just Eq.(\ref{eq3}) obtained in Sec. \ref{sec1}. Since the Haldane model has a quantized Hall conductance, $H_{f}$ also has such Hall conductance $\sigma_{H}=\frac{e^{2}}{h}$ and the edge-state is described by Eq.(\ref{eq9}) or Eq.(\ref{eq10}). The topological properties of $H_{f}$ will be reliably captured by Eq.(\ref{eq7}) or Eq.(\ref{eq8}), as well.

\subsection{The nature of two insulating states}
Since the $f$ fermion is always insulating and has the quantized Hall conductance, the ground-state of the whole system depends on the Ising field part Eq.(\ref{eq3}). From the discussion in previous section, we know the Ising field part can appear in the ferromagnetic, paramagnetic and quantum critical state. In the case of ferromagnetic state $\langle \hat{\tau}^{x} \rangle\neq0$, $c_{i}\simeq \langle\hat{\tau}^{x}\rangle f_{i}$ and the physical electron behaves as the $f$ fermion, which can be identified as the usual Chern insulator as what has been done in Sec. \ref{sec1}. However, if the Ising field is paramagnetic ($\langle\hat{\tau}^{x}\rangle=0$), the physical electrons will not be a well-defined quasiparticle but the $f$ fermion and Ising field will be useful quasiparticle excitations. In this case, the $Z_{2}$ gauge structure is meaningful since the elementary excitations are the $f$ fermion and the Ising field, which both carry the $Z_{2}$ gauge charge. Therefore, we may identify this state as a $Z_{2}$ fractionalized state. When considering that such $Z_{2}$ fractionalized state is still insulating and has the quantized Hall conductance, we may call it the $Z_{2}$ fractionalized Chern insulator. We note that the paramagnetic state corresponds to the deconfined state of the original gauge field and in this case fractionalized excitations,e.g. $f$ fermion and $\hat{\tau}^{x}$, are meaningful and can be choose as the real quasiparticle. Instead, the ferromagnetic state is dual to the confined state of the $Z_{2}$ gauge theory and only physical electron $c$ itself is the real quasiparticle excitations. Therefore, the dual Hamiltonian can reproduce correct results as expected from the original model. When the Ising field part is quantum critical, the system will approach its quantum transition point and the critical behaviors will be determined by the Ising field part of Eq. (\ref{eq4}), which means that the transition from the usual Chern insulator to the $Z_{2}$ fractionalized Chern insulator falls into the classic $3D$ Ising universal class. (The critical exponents for the $3D$ Ising universal class are $\alpha=0.11,\beta=0.32,\gamma=1.24,\delta=4.9,\nu=0.63,\eta=0.04,z=1.$)

It is interesting to see that the two insulating states (the Chern insulator and the $Z_{2}$ fractionalized Chern insulator) discussed above are rather different but they cannot be distinguished by Landau symmetry-breaking theory and the usual topological quantum number such as the (first) Chern number. Therefore, one has to find other methods to identified these two phases when more general and complex models are encountered. A straightforward idea is to check the Green's function of physical electrons in bulk but it does not work since they are both insulating and their Green's function of physical electrons are both gapped. However, the single particle Green's function of physical electrons in the edge may solve this issue since the Chern insulator will have a gapless edge-state while the $Z_{2}$ fractionalized Chern insulator only has a gapped one. We will visit this important issue in the following subsection.

Before proceeding, we should emphasize that the charge of original electrons $c$ is only carried by the $f$ fermion since $\hat{\tau}^{x}$ is a real operator and cannot carry the electromagnetic $U(1)$ charge. Therefore, the quantized Hall conductance $\sigma_{H}=\frac{e^{2}}{h}$ obtained from $H_{f}$ is the true one.
\subsection{The edge-state of the Chern insulator and the $Z_{2}$ fractionalized Chern insulator}

The single particle Green's function for the physical electron is defined by $G(x,t)=-i\langle Tc(x,t)c^{\dag}(0,0)\rangle=-i\langle Tf(x,t)f^{\dag}(0,0)\hat{\tau}^{x}(x,t)\hat{\tau}^{x}(0,0)\rangle$. Since the $f$ fermion and Ising field $\hat{\tau}^{x}$ are decoupled as can be seen in Eq.(\ref{eq12}), one has $G(x,t)=-i\langle Tf(x,t)f^{\dag}(0,0)\rangle\langle T\hat{\tau}^{x}(x,t)\hat{\tau}^{x}(0,0)\rangle$. Recalling that the edge-state for $f$ fermion is described by Eq.(\ref{eq10}) while the low-energy effective description of the Ising fields is the $\varphi^{4}$ theory, the Green's function for the physical electron in the edge will behave as $G(x,t)\propto\frac{1}{x-vt}\times G_{\varphi}^{2}$. $G_{\varphi}^{2}$ is the Green's function of 1+1D $\varphi^{4}$ theory which may describe the Ising field excitation near the edge. When the system is in the $Z_{2}$ fractionalized Chern insulating state (the Ising field is in its paramagnetic phase), $G_{\varphi}^{2}$ should show a gap, which leads to a gap for physical electrons and $G(x,t)$ decays out for long time and distance. In contrast, when one considers the usual Chern insulating state, the Ising field is ferromagnetically ordered in this case, thus we may have $G_{\varphi}^{2}\propto constant$, which means that the physical electrons will have the same behaviors as the $f$ fermion $\left(G(x,t)\propto\frac{1}{x-vt}\right)$.

Interestingly, if the Ising field is critical (this case is realized when the bulk is at its quantum transition point.), the Green's function of physical electrons in the edge will behave as
\begin{eqnarray}
G(x,t)\propto\frac{1}{x-vt}\times\frac{1}{[x^{2}-(ct)^{2}]^{\eta/2}}\label{eq13}
\end{eqnarray}
with $G_{\varphi}^{2}\propto[x^{2}-(ct)^{2}]^{-\eta}$. ($\eta=1/4$ denotes the anomalous dimension of the 2D classical Ising model and $c$ is the edge-velocity for the Ising field.) In the long distance limit, we have
$G(x)\propto x^{-5/4}$, which decays more rapidly than the free electron $G(x)\propto x^{-1}$. The corresponding local density of states is $N(\omega)\propto|\omega|^{1/4}$. In some sense, such explicit Luttinger liquid behaviors, which could be measured by local differential conductance in scanning tunneling microscopy (STM), may be used to locate the exact position of quantum critical point of the system, although the bulk critical fluctuations will dominate in the case.

We should emphasize that although the single electron excitation is gapped in the edge-state of $Z_{2}$ fractionalized Chern insulator, the static and dynamical density-density correlations are still gapless in those edge-states as what have been shown in Ref.[\onlinecite{Ruegg2012}].\cite{Ruegg2013} This point can be seen as follows. The physical static density-density (charge) correlation function in the mean-field approximation reads $\langle c^{\dag}(x) c(x)c^{\dag}(0) c(0)\rangle\approx\langle f^{\dag}(x) f(x)f^{\dag}(0) f(0)\rangle$. Since $f$ fermion is gapless in the edge, the static density-density correlation is expected
to be gapless. For models including the spin degree of freedom, (Those models will be presented in next section.) the static spin-spin correlation will also show the gapless behavior in the edge-state of $Z_{2}$ fractionalized states. For the case of the dynamical two-particle correlation function, e.g. the dynamical density-density correlation $\langle c^{\dag}(x,t) c(x,t)c^{\dag}(0,0) c(0,0)\rangle\approx\langle f^{\dag}(x,t) f(x,t)f^{\dag}(0,0) f(0,0)\rangle$, should also be gapless.

After all, the edge state of physical electrons clearly show distinct behaviors in the Chern insulating and the $Z_{2}$ fractionalized Chern insulating states. Therefore, one can check the behaviors of the single electron excitation in edge-state to identify those two insulating states in generical models.

\section{Exactly soluble models with spin degree of freedom}\label{sec3}

\subsection{Extension to include spin degree of freedom}
Since most real models have the spin degree of freedom, it is crucial to include this element in our model. A simplest extension is to double the Haldane model part, which means electrons carrying different spin-flavor are not interacting at all.
\begin{eqnarray}
&&H=H_{I}+H_{c},\nonumber\\
&&H_{I}=-h\sum_{\langle ij\rangle}\hat{\sigma}^{z}_{ij}-J\sum_{i}(-1)^{c^{\dag}_{i}c_{i}}\prod_{j=i+\delta_{a}}\hat{\sigma}^{x}_{ij}\nonumber\\
&&-W\sum_{i}\prod_{j\in hexagon}\hat{\sigma}^{z}_{ij},\nonumber\\
&&H_{c}=-t\sum_{\langle ij\rangle \sigma}c_{i\sigma}^{\dag}\hat{\sigma}^{z}_{ij}c_{j\sigma}-t'\sum_{\langle\langle ij\rangle\rangle\sigma}e^{i\varphi_{ij}}c_{i\sigma}^{\dag}\hat{\sigma}^{z}_{il}\hat{\sigma}^{z}_{lj}c_{j\sigma}.\label{eq14}
\end{eqnarray}
Since electrons with different spins do not interact, properties of this model is identical to Eq. (\ref{eq1}) expect for doubled quantized Hall conductance and two chiral edge-states. For completeness, we also write down the dual Hamiltonian, which can be readily derived in terms of the same treatment in last section.
\begin{eqnarray}
&&H=H_{\tau}+H_{f},\nonumber\\
&&H_{\tau}=-h\sum_{\langle ij\rangle}\hat{\tau}^{x}_{i}\hat{\tau}^{x}_{j}-J\sum_{i}\hat{\tau}^{z}_{i},\nonumber\\
&&H_{f}=-t\sum_{\langle ij\rangle\sigma}f_{i\sigma}^{\dag}f_{j\sigma}-t'\sum_{\langle\langle ij\rangle\rangle\sigma}e^{i\varphi_{ij}}f_{i\sigma}^{\dag}f_{j\sigma}. \label{eq15}
\end{eqnarray}

\subsection{$Z_{2}$ fractionalized topological insulator}
The next extension for including the spin degree of freedom is to replace the Haldane model part by the well-known Kane-Mele model,\cite{Kane2006}
\begin{eqnarray}
&&H=H_{I}+H_{c},\nonumber\\
&&H_{I}=-h\sum_{\langle ij\rangle}\hat{\sigma}^{z}_{ij}-J\sum_{i}(-1)^{c^{\dag}_{i}c_{i}}\prod_{j=i+\delta_{a}}\hat{\sigma}^{x}_{ij}\nonumber\\
&&-W\sum_{i}\prod_{j\in hexagon}\hat{\sigma}^{z}_{ij},\nonumber\\
&&H_{c}=-t\sum_{\langle ij\rangle \sigma}c_{i\sigma}^{\dag}\hat{\sigma}^{z}_{ij}c_{j\sigma}-t'\sum_{\langle\langle ij\rangle\rangle\sigma}\sigma e^{i\varphi_{ij}}c_{i\sigma}^{\dag}\hat{\sigma}^{z}_{il}\hat{\sigma}^{z}_{lj}c_{j\sigma}\label{eq16}
\end{eqnarray}
where the next-nearest neighbor hopping term is spin-dependent, which mimics the spin-orbit coupling effect in the real materials.
The corresponding dual Hamiltonian reads
\begin{eqnarray}
&&H=H_{\tau}+H_{f},\nonumber\\
&&H_{\tau}=-h\sum_{\langle ij\rangle}\hat{\tau}^{x}_{i}\hat{\tau}^{x}_{j}-J\sum_{i}\hat{\tau}^{z}_{i},\nonumber\\
&&H_{f}=-t\sum_{\langle ij\rangle\sigma}f_{i\sigma}^{\dag}f_{j\sigma}-t'\sum_{\langle\langle ij\rangle\rangle\sigma}\sigma e^{i\varphi_{ij}}f_{i\sigma}^{\dag}f_{j\sigma}. \label{eq17}
\end{eqnarray}
For this model, we expect that the system can show two distinct phases, one is the usual topological insulator (quantum spin Hall insulator)($\langle\hat{\tau}^{x}_{i}\rangle\neq0$) and the other is the $Z_{2}$ fractionalized topological insulator ($\langle\hat{\tau}^{x}_{i}\rangle=0$). These two phases will have no quantized charge Hall conductance ($\sigma_{H}=\sigma_{H}^{\uparrow}+\sigma_{H}^{\downarrow}=0$) but show quantized spin Hall conductance ($\sigma_{SH}=\frac{\hbar}{2e}(\sigma_{H}^{\uparrow}-\sigma_{H}^{\downarrow})=\frac{e}{2\pi}$). For the $Z_{2}$ fractionalized topological insulator, its low-energy properties are
described by the same pure gauge-field theory as what has been discussed in previous section
\begin{eqnarray}
\tilde{H}_{I}=-\widetilde{h}\sum_{\langle ij\rangle}\hat{\sigma}^{z}_{ij}-\widetilde{J}\sum_{i}\prod_{j=i+\delta_{a}}\hat{\sigma}^{x}_{ij}-\widetilde{W}\sum_{i}\prod_{j\in hexagon}\sigma^{z}_{ij}.\nonumber
\end{eqnarray}

For their edge-states, the usual topological insulating states has the helical edge-state
$H_{edge}=\int dx[\psi^{\dag}_{\uparrow}(-iv\partial_{x})\psi_{\uparrow}-\psi^{\dag}_{\downarrow}(-iv\partial_{x})\psi_{\downarrow}]$ and this gapless edge-state is protected by the time-reversal symmetry. For the case of the
$Z_{2}$ fractionalized topological insulator, the single particle excitation of physical electrons in the edge-state is gapped due to the massive Ising spin field $\hat{\tau}^{x}$.

Besides, when the system is critical, the edge-state for the physical electrons
has the behavior described by Eq.(\ref{eq13}) with two spin-flavors. One may write down the corresponding Luttinger liquid Hamiltonian for such critical case\cite{Giamarchi}
\begin{eqnarray}
H_{edge}=\frac{1}{2\pi}\int dx[uK(\partial_{x}\theta)^{2}+\frac{u}{K}(\partial_{x}\phi)^{2}]\nonumber
\end{eqnarray}
where $u$ denotes the non-universal velocity and $K$ is the Luttinger parameter. The fermion (electron) operator is defined by $\psi_{\sigma}(x)\propto e^{-i(\phi(x)-\sigma\theta(x))}$. One may find that the Luttinger parameter $K=2$ or $K=1/2$ both can reproduce power-law behavior of the electron Green's function. However, generally, $K>1$ means the electrons are very likely to be pairing and considering no such instability exists in the original model, we conclude one should choose $K=1/2$ and the local density of state for the physical electrons is $N(\omega)\propto|\omega|^{1/4}$. The related momentum distribution function of physical electrons is $n(k)\sim$constant$+|k|^{1/4}sgn(k)$ and one clearly sees no jump appears near the Fermi momentum ($k=0$). The density-density correlation in this edge-state is easy to find and it decays as $\frac{1}{x}$ in the large distance limit while the free electron case gives a $\frac{1}{x^{2}}$ behavior.

In Ref.[\onlinecite{Levin2009}], authors argued the existence of the fractional topological insulators, which can be realized by requiring that the two spin species in the usual topological insulator each form fractional quantum
Hall states. The name ``fractional" topological insulators results from the natural extension of the usual fractional quantum Hall effect and its corresponding topological order is expected to be the similar one in the fractional quantum Hall states. For our $Z_{2}$ fractionalized topological insulator, we have the $Z_{2}$ topological order and do not involve the fractional quantum Hall-like objects. Thus, we think that the $Z_{2}$ fractionalized topological insulator is different from the fractional topological insulators proposed in Ref.[\onlinecite{Levin2009}].

\subsection{Comparing the $Z_{2}$ fractionalized topological insulator to the QSH$^{\ast}$ state}

In Ref.[\onlinecite{Ruegg2012}], an unusual $Z_{2}$ gapped states (fractionalized quantum spin Hall state ($QSH^{\ast}$)) is discovered by using the $Z_{2}$ slave-spin mean-field approximation on the honeycomb lattice.
Such state is stable when local repulsive energy and spin-orbit coupling are both strong and it has quantized spin Hall conductance as the $Z_{2}$ fractionalized topological insulator studied in the last subsection.
In the low-energy limit, $QSH^{\ast}$ is described by a pure $Z_{2}$ lattice gauge theory while the $Z_{2}$ fractionalized topological insulator can also be described the pure $Z_{2}$ gauge theory. (One may simply discard the electrons $c$ in Eq.(\ref{eq15}) since they cannot contribute singular correction to the $Z_{2}$ gauge fields.) Furthermore, the single electron excitations in edge-states for $QSH^{\ast}$ and the $Z_{2}$ fractionalized topological insulator are gapped.
Therefore, we may think the $Z_{2}$ fractionalized topological insulator found here is similar to the QSH$^{\ast}$ state in Ref.[\onlinecite{Ruegg2012}]. In some sense, our model may provide a possible realization for QSH$^{\ast}$ state in exactly soluble models.

\section{Extensions and Discussions} \label{sec4}

\subsection{Relation to the real microscopic models}
The models Eqs.(\ref{eq1}),(\ref{eq14}),(\ref{eq16}) introduced in the present paper have the $Z_{2}$ gauge field part, which is rather artificial and dose not appear in many real models for condensed matter physics.
Therefore, it is important to find some real and simple models which can provide the similar states as what have been found in our main text.

The first model is the topological Hubbard model on the honeycomb lattice at half-filling,\cite{He2011}
\begin{eqnarray}
H&&=-t\sum_{\langle ij\rangle\sigma}(c_{i\sigma}^{\dag}c_{j\sigma}+h.c.)-t'\sum_{\langle\langle ij\rangle\rangle\sigma}e^{i\varphi_{ij}}c_{i\sigma}^{\dag}c_{j\sigma}\nonumber\\
&&+\frac{U}{2}\sum_{i}(n_{i}-1)^{2},\label{eq18}
\end{eqnarray}
where $n_{i}=\sum_{\sigma}c_{i\sigma}^{\dag}c_{i\sigma}$, $U$ is the onsite Coulomb energy between electrons on the same site and $t$ is the hopping energy between nearest-neighbor sites. Since we are interested in the case of half-filling, the chemical potential has been set to zero.

Then in terms of $Z_{2}$ slave-spin representation,\cite{deMedici,Ruegg} the physical electron $c_{\sigma}$ is fractionalized into a new slave-fermion $f_{\sigma}$ and a slave-spin $\tau^{x}$ as
$c_{i\sigma}=f_{i\sigma}\tau_{i}^{x}$ with a constraint $\tau_{i}^{z}+1=2(n_{i}-1)^{2}$ enforced in every site. Under this representation, the original Hamiltonian can be rewritten as
\begin{eqnarray}
H&&=-t\sum_{\langle ij\rangle \sigma}(\tau_{i}^{x}\tau_{j}^{x}f_{i\sigma}^{\dag}f_{j\sigma}+h.c.)-t'\sum_{\langle\langle ij\rangle\rangle\sigma}e^{i\varphi_{ij}}\tau_{i}^{x}\tau_{j}^{x}c_{i\sigma}^{\dag}c_{j\sigma}\nonumber\\
&&+\frac{U}{4}\sum_{i}(\tau_{i}^{z}+1)\label{eq19}
\end{eqnarray}
where $n_{i}=n_{i}^{f}=\sum_{\sigma}f_{i\sigma}^{\dag}f_{i\sigma}$. Obviously, a $Z_{2}$ local gauge symmetry is left in this representation and the corresponding low-energy effective theory should respect this. The mentioned gauge structure can be seen if $f_{i\sigma}^{(\dag)}\rightarrow \epsilon_{i}f_{i\sigma}^{(\dag)}$ and $\tau_{i}^{x}\rightarrow\epsilon_{i}\tau_{i}^{x}$ with $\epsilon_{i}=\pm1$ while the whole Hamiltonian $H$ is invariant under this $Z_{2}$ gauge transformation.

When one utilizes the mean-field approximation, the resulting mean-field Hamiltonian will be
\begin{eqnarray}
&&H_{f}=-\sum_{\langle ij\rangle\sigma}(\tilde{t}_{ij}f_{i\sigma}^{\dag}f_{j\sigma}+h.c.)-\sum_{\langle\langle ij\rangle\rangle\sigma}\tilde{t}'_{ij}e^{i\varphi_{ij}}c_{i\sigma}^{\dag}c_{j\sigma},\label{eq20}\\
&&H_{I}=-\sum_{\langle ij\rangle\sigma}(J_{ij}\tau_{i}^{x}\tau_{j}^{x}+h.c.)-\sum_{\langle\langle ij\rangle\rangle}J'_{ij}\tau_{i}^{x}\tau_{j}^{x}+\frac{U}{4}\sum_{i}\tau_{i}^{z},\label{eq21}
\end{eqnarray}
where we have defined $\tilde{t}_{ij}=t\langle \tau_{i}^{x}\tau_{j}^{x}\rangle$,$\tilde{t}'_{ij}=t'\langle \tau_{i}^{x}\tau_{j}^{x}\rangle$, $J_{ij}=t\sum_{\sigma}\langle f_{i\sigma}^{\dag}f_{j\sigma}\rangle$, $J'_{ij}=t'e^{i\varphi_{ij}}\sum_{\sigma}\langle f_{i\sigma}^{\dag}f_{j\sigma}\rangle$. The decoupled Hamiltonian $H_{I}$ is an extended quantum Ising model in transverse field and $H_{f}$ describes $f$ fermions in the honeycomb lattice. Here, all the Lagrange multipliers have been set to zero, provided only non-magnetic solutions are involved and a half-filling case is considered.\cite{Ruegg}

Now, comparing Eqs.(\ref{eq20}),(\ref{eq21}) to Eq.(\ref{eq15}), the mean-field Hamiltonian is able to capture the basic features of the finely tuned exactly soluble models except for the extra trivial next-nearest-neighbor coupling term in the quantum Ising model part. However, one should realize that the mean-field treatment is a biased method and the original model may not have to support the putative $Z_{2}$ fractionalized states.

If one wants to go beyond the mean-field approximation, the intrinsic $Z_{2}$ gauge structure should be respected and the above mean-field Hamiltonian could be modified into the following ones
\begin{eqnarray}
&&\tilde{H}_{f}=-\sum_{\langle ij\rangle\sigma}(\tilde{t}_{ij}f_{i\sigma}^{\dag}\hat{\varrho}^{z}_{ij}f_{j\sigma}+h.c.)-\sum_{\langle\langle ij\rangle\rangle\sigma}\tilde{t}'_{ij}e^{i\varphi_{ij}}c_{i\sigma}^{\dag}\hat{\varrho}^{z}_{il}\hat{\varrho}^{z}_{lj}c_{j\sigma},\nonumber\\
&&\tilde{H}_{I}=-\sum_{\langle ij\rangle\sigma}(J_{ij}\tau_{i}^{x}\hat{\varrho}^{z}_{ij}\tau_{j}^{x}+h.c.)-\sum_{\langle\langle ij\rangle\rangle}J'_{ij}\tau_{i}^{x}\hat{\varrho}^{z}_{il}\hat{\varrho}^{z}_{lj}\tau_{j}^{x}\nonumber\\
&&\hspace{3cm}+\frac{U}{4}\sum_{i}\tau_{i}^{z}\nonumber\\
&&H_{Z_{2}}=-h\sum_{\langle ij\rangle}\hat{\varrho}^{x}_{ij}-G\sum_{i}\prod_{j\in hexagon}\hat{\varrho}^{z}_{ij}
\end{eqnarray}
where $\hat{\varrho}^{z}_{ij}$ denotes the dynamical $Z_{2}$ gauge-field, which is introduced to enforce the desirable intrinsic $Z_{2}$ gauge structure. For convenience, we have added the $Z_{2}$ gauge-field part $H_{Z_{2}}$ and
the gauge transformation operator is defined by $\hat{G}_{i}=(-1)^{\frac{1}{2}\tau_{i}^{z}-\frac{1}{2}+\sum_{\sigma}f_{i\sigma}^{\dag}f_{i\sigma}}\prod_{j=i+\delta_{a}}\hat{\varrho}^{x}_{ij}$. If we focus on the fractionalized states, the $\tilde{H}_{I}$ could be dropped out and the gauge transformation operator is now read as $\hat{G}_{i}=(-1)^{\sum_{\sigma}f_{i\sigma}^{\dag}f_{i\sigma}}\prod_{j=i+\delta_{a}}\hat{\varrho}^{x}_{ij}$. Because $\hat{G}_{i}^{2}=1$ and $[\hat{G}_{i},\tilde{H}_{f}+H_{Z_{2}}]=0$, we may set $\hat{G}_{i}=1$ for physical states. Then, it is straightforward to see that if a $Z_{2}$ charge ($\prod_{j=i+\delta_{a}}\hat{\varrho}^{x}_{ij}=-1$) is created at site $i$, this site must be occupied by a slave-fermion $f_{\sigma}$ so as to fulfill the constraint $\hat{G}_{i}=1$. In contrast, if no $Z_{2}$ charge appears, the corresponding site should not have a slave-fermion $f_{\sigma}$. Therefore, we may say that the $Z_{2}$ charge is bound to the slave-fermion and the mutual statistics of slave-fermion and the $Z_{2}$ vortex ($\prod_{j\in hexagon}\hat{\varrho}^{z}_{ij}=-1$) is the semionic
statistics. (Recalling that the semionic mutual statistics between the $Z_{2}$ charge and the $Z_{2}$ vortex is discussed in Sec.\ref{sec2}.)
More detailed properties on similar model has been investigated in Ref.[\onlinecite{Ruegg2012}] and we refer interested reader to their original paper.

In addition, when one wants to describe the $Z_{2}$ fractionalized topological insulator in Eq.(\ref{eq16}), the simplest model will be given by
\begin{eqnarray}
H&&=-t\sum_{\langle ij\rangle\sigma}(c_{i\sigma}^{\dag}c_{j\sigma}+h.c.)-t'\sum_{\langle\langle ij\rangle\rangle\sigma}\sigma e^{i\varphi_{ij}}c_{i\sigma}^{\dag}c_{j\sigma}\nonumber\\
&&+\frac{U}{2}\sum_{i}(n_{i}-1)^{2}.\label{eq22}
\end{eqnarray}

\subsection{The exactly soluble model on the square lattice}
In the square lattice, the gauge-field part is now read as
\begin{eqnarray}
H_{I}&&=-h\sum_{\langle ij\rangle}\hat{\sigma}^{z}_{ij}-J\sum_{i}\prod_{j=i\pm \hat{x},i\pm \hat{y}}\hat{\sigma}^{x}_{ij}\nonumber\\
&&-W\sum_{i}\prod_{j\in plaquett}\sigma^{z}_{ij}\label{eq23}
\end{eqnarray}
where the $Z_{2}$ gauge transformation operator is defined as $\hat{G}_{i}=\prod_{j\in plaquett}\hat{\sigma}^{z}_{ij}$.

For the electron part, a simple and feasible choice is \cite{Ng}
\begin{eqnarray}
H_{c}&&=t\sum_{i_{x}}(c_{i_{x}\uparrow}^{\dag}\hat{\sigma}^{z}_{i_{x}i_{x}+1}c_{i_{x}+1\downarrow}-c_{i_{x}\uparrow}^{\dag}\hat{\sigma}^{z}_{i_{x}i_{x}-1}c_{i_{x}-1\downarrow})\nonumber\\
&&+it\sum_{i_{y}}(c_{i_{y}\uparrow}^{\dag}\hat{\sigma}^{z}_{i_{y}i_{y}+1}c_{i_{y}+1\downarrow}-c_{i_{y}\uparrow}^{\dag}\hat{\sigma}^{z}_{i_{y}i_{y}-1}c_{i_{y}-1\downarrow})\nonumber\\
&&+H.c.+\sum_{i}m_{z}(c_{i\uparrow}^{\dag}c_{i\uparrow}-c_{i\downarrow}^{\dag}c_{i\downarrow})\nonumber\\
&&+t'\sum_{\langle ij\rangle}(c_{i\uparrow}^{\dag}c_{j\uparrow}-c_{i\downarrow}^{\dag}c_{j\downarrow})\label{eq24}
\end{eqnarray}
where $i=(i_{x},i_{y})$ and only nearest-neighbor-hopping is involved. $m_{z}$ denotes a tunable parameter which may result from the usual ferromagnetic order.  We also require the implicit half-filling condition for this model.
Following the treatment in the main text, the electron part will be dual to a free auxiliary fermions part, whose Hamiltonian reads
\begin{eqnarray}
H_{f}=-\sum_{k}f_{k\sigma}^{\dag}[\vec{d}(k)\cdot\hat{\varrho}]_{\sigma\sigma'}f_{k\sigma} \nonumber
\end{eqnarray}
with $\hat{\varrho}_{x},\hat{\varrho}_{y},\hat{\varrho}_{z}$ being the usual Pauli matrix acting on the spin space.
We have defined $\textbf{d}(k)=(2t\sin k_{y},2t\sin k_{x},-m_{z}+2t'\cos k_{x}+2t'\cos k_{y})$. As what has been done in the Haldane model, Hall conductance of this model is calculated by
$\sigma_{H}=C_{1}\frac{e^{2}}{h}$ with $C_{1}=-\frac{1}{4\pi}\int dk_{x}dk_{y}\hat{\textbf{d}}\cdot(\frac{\partial\hat{\textbf{d}}}{\partial k_{x}}\times \frac{\partial\hat{\textbf{d}}}{\partial k_{y}})$.
One finds $C_{1}=\frac{|m_{z}|}{m_{z}}$ for $0<|m_{z}|<4t'$ while $C_{1}$ vanishes in other cases. Such Hall conductance can also be obtained by expanding the above Hamiltonian around four high-symmetry points of the energy band.
($(0,0),(\pi,0),(0,\pi),(\pi,\pi)$)
\begin{eqnarray}
S=\int d^{2}xd\tau\sum_{a}[\bar{\psi}_{a}(\gamma_{\mu}(\partial_{\mu}-ieA_{\mu})+m_{a})\psi_{a}]\nonumber
\end{eqnarray}
where the effective mass has been defined as $m_{a=(0,0)}=4t'-m_{2}$ and $m_{a=(\pi,\pi)}=4t'+m_{2}$ when $4t'$ close to $|m_{2}|$. If $m_{2}\simeq0$, $m_{a=(\pi,0)}=m_{a=(0,\pi)}=m_{2}$.
We note the effective Dirac theory only has two-flavor, thus $C_{1}$ can take only $0,\pm1$. Other properties are all similar to the case of the previous discussion and we will not present them here.

\subsection{The model for the topological superconductor}
We have discussed the usual/fractionalized Chern/topological insulators in the previous sections and one may also expect topological superconductor-like states may be described by similar models in the main text.
However, the exactly model is unable to construct because the superconducting pairing is a many-body effect and we have to rely on some mean-field treatment. A straight way to obtain a topological superconductor-like state is to introduced attractive local interaction $H_{U}=-U\sum_{i}c_{i\uparrow}^{\dag}c_{i\uparrow}c_{i\downarrow}^{\dag}c_{i\downarrow}$ into Eq.(\ref{eq24}). According to Ref.[\onlinecite{Ng}[, assuming a s-wave paring and solving the corresponding mean-field self-consistent equation for the superconducting order parameter, one can obtain an effective $p+ip$ chiral superconducting phase with non-zero Chern number $C_{1}$. More details can be found in Ref.[\onlinecite{Ng}]. Then, combing the electron part with the $Z_{2}$ gauge-field, we expect there exists a transition from the chiral superconducting phase to the $Z_{2}$ fractionalized chiral superconductor.

\subsection{Relation to the fractional Chern insulators}
It is also interesting to discuss the relation of $Z_{2}$ fractionalized Chern insulators in our paper to the fractional Chern insulators in Refs.[\onlinecite{Neupert,Mei,Sun,Regnault,Sheng2011}].
We note that in their model, the filling of electrons is fractional, thus the system is metallic without introducing interactions while our case is insulating since half-filling is explicitly assumed.
To achieve the fractional quantum Hall-like states, the noninteracting part of the Hamiltonian is finely tuned into nearly-dispersionless (flat) as the case for the flat Landau levels in the classic fractional quantum Hall effect. Then, including the effect of certain proper interactions leads to fractional quantum Hall-like states found in their sophisticated numerical calculation.
For our models, the Landau levels-like element is not involved, thus we do not require the flat band condition. The fractionalized insulating state in our models will have $Z_{2}$ topological order in contrast to the $m$-fold ($m=1/\nu$ with $\nu$ being the particular fractional filling) topological degeneracy in the fractional quantum Hall-like states. In other words, on a torus, $Z_{2}$ fractionalized Chern insulators discussed in the main text have
four degenerate ground-states while $m$ degenerate ground-states appear in the fractional Chern insulators.
Therefore, these two kinds of fractionalized states are quite different in their nature and it seems still a challenge to construct even artificial exactly soluble models for such fractionalized Chern insulators.

\subsection{Relation to the orthogonal metals}
The orthogonal metals are metallic in their original definition but it can also extend to gapped phases.\cite{Nandkishore,Zhong2012e,Zhong2012,Zhong2013b} When such extension is made, one expects a fractionalized state and a non-fractionalized counterpart. These two phases should have the same bulk thermal and transport properties. From the discussion in the previous sections, we realize that the $Z_{2}$ fractionalized Chern/topological insulators in exactly soluble correlated models can correspond to the fractionalized state and the usual Chern/topological insulators relate to the non-fractionalized ones in orthogonal metal-like states. In the sense of orthogonal metals, the $Z_{2}$ fractionalized Chern/topological insulators can also be dubbed as orthogonal Chern/topological insulators since bulk thermal (gapped excitations) and transport properties (quantized charge/spin Hall conductance) are identical with single particle excitation different. An extra element for the present orthogonal Chern/topological insulators, which is not noticed in original studies,\cite{Nandkishore,Zhong2012e,Zhong2012,Zhong2013b} comes from their edge-states for physical electrons as shown in the cases for the square and honeycomb lattices.

\subsection{Relation to the realistic materials or systems}
It is important to find which realistic materials may realize the proposed fractionalized Chern/topological insulators.
We think that the fractionalized Chern/topological insulator we proposed may
be found in systems with strong spin-orbit coupling and strong electron-electron interaction, however, the current topological insulator materials is well described by single electron approximation. Thus, usual solid state materials are not good candidates and we have to suspect that the versatile cold atom system may have an opportunity to realize the proposed fractionalized Chern/topological insulator in future.

\section{conclusion} \label{sec5}
In summary, we have propose certain exactly soluble models which support $Z_{2}$ fractionalized Chern/topological insulators besides the usual Chern/topological insulating states. The bulk behaviors of physical electrons are similar in these two states but the edge-states of physical electrons have rather different behaviors, which provides a definite signature to identify the fractionalized states from the non-fractionalized ones. Besides, the transition from the usual Chern insulator to the $Z_{2}$ fractionalized Chern insulator is found to fall into the usual 3D Ising universal class.

Moreover, we have inspected relations to the QSH$^{\ast}$ in Ref.\onlinecite{Ruegg2012} and have made an extension to the case of the square lattice. For the specific case of the square lattice, we expect that a chiral topological superconducting phase and its $Z_{2}$ fractionalized version may appear when the attractive local interaction is introduced. We also provide intimate link of our exactly soluble models to various more real lattice models without intrinsic degree of freedom for gauge fields. In terms of the $Z_{2}$ slave-spin mean-field approximation,\cite{deMedici,Ruegg} we have suggested that similar fractionalized states found in our present paper may appear in those lattice models. The relation of $Z_{2}$ fractionalized Chern insulators found in the main text to the fractional Chern insulator Refs. [\onlinecite{Neupert,Mei,Sun,Regnault,Sheng2011}] is inspected, which implies that these two kinds of fractionalized states are quite distinct in their nature. We also provide some discussion on the relations to orthogonal metals\cite{Nandkishore,Zhong2012e,Zhong2012,Zhong2013b} and fractional topological insulators.\cite{Levin2009} The present work may be helpful for further studies on the fractional Chern/topological insulator and the related novel strongly correlated quantum phases.

After the completion of this work, we are aware of the work of Maciejko and R\"{u}egg\cite{Ruegg2013}, who study the $Z_{2}$ fractionalized Chern insulators (also called $CI^{\ast}$ in their paper) and have obtained similar conclusion with our results.

\begin{acknowledgments}
We thank A. R\"{u}egg for helpful discussions. The work was supported partly by NSFC, PCSIRT (Grant No. IRT1251), the Program for NCET, the Fundamental Research Funds for the Central Universities and the national program for basic research of China.
\end{acknowledgments}

\appendix
\section{The standard $Z_{2}$ lattice gauge theory model on the honeycomb lattice and its topological order}
The standard $Z_{2}$ lattice gauge theory model on the honeycomb lattice is defined by the following Hamiltonian,
\begin{eqnarray}
H_{Z_{2}}=-h\sum_{\langle ij\rangle}\hat{\sigma}^{x}_{ij}-J\sum_{i}\prod_{j\in hexagon}\hat{\sigma}^{z}_{ij}
\end{eqnarray}
where the first term describes the fluctuation induced by $\hat{\sigma}^{x}_{ij}$ while the second one denotes the potential gained by six Ising field $\hat{\sigma}^{z}_{ij}$ in the same hexagon. One can define the $Z_{2}$ gauge transformation operator $\hat{G}_{i}=\prod_{j=i+\delta_{a}}\hat{\sigma}^{x}_{ij}$ and it is easy to check that $[\hat{G}_{i},H_{Z_{2}}]=0$. Thus, the model $H_{Z_{2}}$ has the wanted $Z_{2}$ gauge invariance and any physical states should be invariant under the operation of $\hat{G}_{i}$.
There are two limit situations which are easy to analyze. First, if $J\gg h$, then the ground-state should have $\prod_{j\in hexagon}\hat{\sigma}^{z}_{ij}=1$ for all hexagons and this is just the deconfined state for the $Z_{2}$ gauge theory. However, when $J\ll h$, one may set all $\hat{\sigma}^{x}_{ij}=1$ to reach the ground-state and the obtained state is the so-called confined state for the $Z_{2}$ gauge theory.

Then, we briefly discuss the topological order of the $Z_{2}$ lattice gauge theory in its deconfined state.\cite{Senthil2000b} Considering a cylindrical geometry and assuming the we work in the deconfined phase, the ground-state is easily obtained
by setting all of $\hat{\sigma}^{z}_{ij}=1$ and this corresponds to the case without threading a $Z_{2}$ vortex (vison)(A $Z_{2}$ vortex is created by setting $\hat{\sigma}^{z}_{ij}=-1$ in certain link while keeping others unchanged.) into the hole of the cylinder.
If we thread a $Z_{2}$ vortex into the hole of the cylinder, the resulting state is also the ground-state. Thus, the
deconfined phase of the $Z_{2}$ gauge theory has two degenerate ground states. Using the same argument, the the deconfined phase of the $Z_{2}$ gauge theory on a
torus will have four degenerate ground states corresponding to the $Z_{2}$ vortex threading or not threading each of the two holes.

The benefit of the above model is that there exists an exact dual transformation which transforms the $Z_{2}$ lattice gauge theory model into a quantum Ising model on the honeycomb lattice. The dual transformation is defined as
\begin{equation}
\prod_{j\in hexagon}\hat{\sigma}^{z}_{ij}=\hat{\tau}^{x}_{i},
\end{equation}
\begin{equation}
\hat{\sigma}^{x}_{ij}=\hat{\tau}^{z}_{i}\hat{\tau}^{z}_{j}.
\end{equation}
Then, the original Hamiltonian is transformed into the following one
\begin{eqnarray}
H=-h\sum_{\langle ij\rangle}\hat{\tau}^{z}_{i}\hat{\tau}^{z}_{j}-J\sum_{i}\hat{\tau}^{x}_{i},
\end{eqnarray}
One may note that this is just the quantum transverse Ising model defined on the honeycomb lattice. Reader may care whether some constraints are needed for the transformed Hamiltonian $H$. It is crucial to see that the $Z_{2}$ gauge transformation operator $\hat{G}_{i}=\prod_{j=i+\delta_{a}}\hat{\sigma}^{x}_{ij}=\prod_{j=i+\delta_{a}}\hat{\tau}^{z}_{j}$ and this should be taken as a constraint in the dual quantum Ising model. Therefore, in general, the resulting quantum transverse Ising model is not free. If we simply neglect the effect of the constraint, the quantum transverse Ising model will have two phase characterized by the Landau local order parameter $\langle\hat{\tau}^{z}_{i}\rangle=0$ (the magnetically disordered phase) and $\langle\hat{\tau}^{z}_{i}\rangle\neq0$ (the ferromagnetic phase). There also exists a second-order quantum phase transition between those two phases when $t/J$ is finely tuned into the critical value. The critical exponents of this quantum model is identical to the classical Ising model in three space dimension. ($\alpha=0.11,\beta=0.32,\gamma=1.24,\delta=4.9,\nu=0.63,\eta=0.04,z=1.$)

\section{Path integral and effective theory for the quantum transverse Ising model}
The quantum Ising model in transverse field is defied as\cite{Sachdev2011}
\begin{equation}
\hat{H}_{I}=-h\sum_{\langle ij\rangle}(\hat{\tau}_{i}^{z}\hat{\tau}_{j}^{z}+h.c.)-J\sum_{i}\hat{\tau}_{i}^{x}
\end{equation}
where a ferromagnetic coupling $h>0$ is assumed and $J$ represents the the transverse external field.

At first glance, one may directly use the coherent state of spin operators in constructing the path integral representation, (One can find a brief but useful introduction to this issue in Ref. [\onlinecite{Sachdev2011}]) however, this will lead to an extra topological Berry phase term and is not easy to utilize practically. An alterative approach is to use the eigenstates of spin operator $\tau^{x}$ or $\tau^{z}$ as the basis for calculation.\cite{Stratt}
One will see this approach is free of the topological Berry phase term and give rise to a rather simple formalism.  Therefore, to construct a useful path integral representation, we will follow Ref. [\onlinecite{Stratt}].

First of all, we consider the orthor-normal basis of $N_{s}$-Ising spins as
\begin{equation}
|\sigma\rangle\equiv|\sigma_{1}\rangle|\sigma_{2}\rangle|\sigma_{2}\rangle\cdot\cdot\cdot|\sigma_{N}\rangle
\end{equation}
with $\sigma_{i}=\pm1$ and define
\begin{equation}
\tau_{i}^{z}|\sigma\rangle=\sigma_{i}|\sigma\rangle,
\end{equation}
\begin{equation}
\tau_{i}^{x}|\sigma\rangle
=|\sigma_{1}\rangle|\sigma_{2}\rangle|\sigma_{3}\rangle\cdot\cdot\cdot|-\sigma_{i}\rangle\cdot\cdot\cdot|\sigma_{N}\rangle.
\end{equation}
Then the partition function $Z=Tr(e^{-\beta \hat{H}})$ can be represented as
\begin{eqnarray}
Z=\sum_{\{\sigma\}=\pm1}\prod_{n=1}^{N}e^{\epsilon h\sum_{\langle ij\rangle}\sigma_{i}(n)\sigma_{j}(n)}\langle\sigma(n+1)|e^{\epsilon J\sum_{i}\tau_{i}^{x}}|\sigma(n)\rangle \nonumber
\end{eqnarray}
where $\epsilon$N=$\beta$.
The calculation of $\langle\sigma(n+1)|e^{\epsilon J\sum_{i}\tau_{i}^{x}}|\sigma(n)\rangle$ is straightforward
by exponentiating the $\tau_{i}^{x}$ matrix and one gets
\begin{eqnarray}
\langle\sigma(n+1)|e^{\epsilon J\sum_{i}\tau_{i}^{x}}|\sigma(n)\rangle&&=\frac{1}{2}(e^{\epsilon J}+e^{-\epsilon J}\sigma_{i}(n)\sigma_{i}(n+1)),\nonumber \\
&&=e^{a\sigma_{i}(n)\sigma_{i}(n+1)+b}
\end{eqnarray}
where $a=\frac{1}{2}[\ln\cosh(\epsilon J)-\ln\sinh(\epsilon J)]$ and $b=\frac{1}{2}[\ln\cosh(\epsilon J)+\ln\sinh(\epsilon J)]$.
Therefore, the resulting path integral formalism for the quantum Ising model in transverse field is
\begin{equation}
Z=\sum_{\{\sigma\}=\pm1}\prod_{n=1}^{N}e^{\epsilon h\sum_{\langle ij\rangle}\sigma_{i}(n)\sigma_{j}(n)+\sum_{i}a\sigma_{i}(n)\sigma_{i}(n+1)+N_{s}b}.
\end{equation}

Further, if one assumes the model is defined in a hyper-cubic lattice in space dimension of d, an effective theory can be derived as
\begin{equation}
Z=\int D\phi \delta(\phi^{2}-1) e^{-\int d\tau d^{d}x \frac{1}{2g}[(\partial_{\tau}\phi)^{2}+c^{2}(\nabla\phi)^{2}]},
\end{equation}
where $\frac{1}{2g}=(\frac{a\epsilon}{a_{0}^{d}})^{\frac{d+1}{2}}$ with $a_{0}$ being the lattice constant and $c^{2}=\frac{ha_{0}^{d-2}}{a\epsilon}$. Moreover, in the effective theory, $\phi$ corresponds to $\tau^{z}$ while $\tau^{x}$ gives the kinetic energy term in imaginary time. Then, the standard $\phi^{4}$ theory is obtained by relaxing the hard constraint $\phi^{2}=1$ while introducing a potential energy term,
\begin{equation}
Z=\int D\phi e^{-\int d\tau d^{d}x [(\partial_{\tau}\phi)^{2}+c^{2}(\nabla\phi)^{2}+r\phi^{2}+u\phi^{4}]},
\end{equation}
where $r,u$ are effective parameters depending on microscopic details.

For the case of the honeycomb lattice, since only the modes near the minimum energy region of the band are involved in low-energy limit, the quantum Ising model on the honeycomb lattice can also be described in terms of the above $\varphi^{4}$ theory in spite of its bipartite feature.

\section{Derivation of Chern-Simon action}
Here, we would like to give a brief derivation of the effective Chern-Simon action Eq. (\ref{eq6}) from the Dirac action Eq. (\ref{eq5}).
First, the Dirac action is written as
\begin{eqnarray}
S=\int d^{2}xd\tau\sum_{a\sigma}[\bar{\psi}_{a\sigma}(\gamma_{\mu}(\partial_{\mu}-ieA_{\mu})+m)\psi_{a\sigma}].
\end{eqnarray}
Then, integrating out Dirac fermions one obtains
\begin{eqnarray}
&& S_{eff}=N\ln Det[\gamma_{\mu}(\partial_{\mu}-ieA_{\mu})+m]\nonumber\\
&& =N Tr\ln[\gamma_{\mu}(\partial_{\mu}-ieA_{\mu})+m]\nonumber\\
&& =N Tr[\ln[\gamma_{\mu}\partial_{\mu}+m]+\ln[1-ie(\gamma_{\mu}\partial_{\mu}+m)^{-1}\gamma_{\mu}A_{\mu}]]\nonumber\\
&& \simeq N \int\frac{d^{3}q}{(2\pi)^{3}}A_{\mu}\Pi_{\mu\nu}(q)A_{\nu}
\end{eqnarray}
where $\Pi_{\mu\nu}(q)=\frac{-e^{2}}{2}\int\frac{d^{3}k}{(2\pi)^{3}}Tr[\frac{ik_{\mu}\gamma_{\mu}-m}{m^{2}+k^{2}}\gamma_{\nu}\frac{i(k_{\mu}+q_{\mu})\gamma_{\mu}-m}{m^{2}+(k+q)^{2}}\gamma_{\mu}]
=\frac{-e^{2}}{2}\int\frac{d^{3}k}{(2\pi)^{3}}[\frac{1}{m^{2}+k^{2}}\frac{1}{m^{2}+(k+q)^{2}}][-imq_{\lambda}Tr(\gamma_{\mu}\gamma_{\nu}\gamma_{\lambda})]+...
\simeq\frac{-e^{2}m}{8\pi|m|}\epsilon^{\mu\nu\lambda}q_{\lambda}$ is calculated at one-loop level. We have also used the identity $Tr(\gamma_{\mu}\gamma_{\nu}\gamma_{\lambda})=2i\epsilon^{\mu\nu\lambda}$ with $\gamma_{0}=\tau_{z}$, $\gamma_{1}=\tau_{x}$ and $\gamma_{2}=\tau_{y}$ while $\int\frac{d^{3}k}{(2\pi)^{3}}[\frac{1}{m^{2}+k^{2}}\frac{1}{m^{2}+(k+q)^{2}}]=\frac{\arcsin\left(\frac{|q|}{\sqrt{q^{2}+4m^{2}}}\right)}{4\pi|q|}\simeq\frac{1}{8\pi|m|}$ for $|q|\ll|m|$. Therefore, the effective Chern-Simon action Eq. (\ref{eq6}) is obtained as
\begin{eqnarray}
S_{eff}&&=\int\frac{d^{3}q}{(2\pi)^{3}} N A_{\mu}\frac{-e^{2}m}{8\pi|m|}\epsilon^{\mu\nu\lambda}q_{\lambda}A_{\nu}\nonumber\\
&&=\int d^{2}xd\tau[N e^{2}\frac{-i m}{8\pi|m|}\epsilon^{\mu\nu\lambda}A_{\mu}\partial_{\nu}A_{\lambda}].
\end{eqnarray}

\end{document}